\documentclass[twoside]{article}
\usepackage{latexsym}
\usepackage{amsmath}
\usepackage{amsfonts}
\usepackage{amssymb}
\usepackage{times}
\usepackage{epsfig}
\newcommand{\JJ}{\mathbin{\raisebox{0.25ex}{$\footnotesize
                       \rm\vphantom{I}%
                       \_\hskip -0.25em\_%
                       \vrule width 0.6pt$}}}           

\newcommand{\cl}{C \kern -0.1em \ell}     

\newcommand{\ut}[1]{{\setbox0=\hbox{$#1$}\mathsurround=0pt
       \rlap{\raisebox{-0.8\dp0}{\raisebox{-0.8ex}
       {\kern -0.15ex\hbox{$\tiny\sim$}\kern 0.15ex}}}#1}}
\newcommand{\uti}[1]{{\setbox0=\hbox{$#1$}\mathsurround=0pt
       \rlap{\raisebox{-0.8\dp0}{\raisebox{-0.8ex}
       {\kern -0.3ex\hbox{$\tiny\sim$}\kern 0.3ex}}}#1}}

\newdimen\arrayruleHwidth                 
     \makeatletter
     \def\Hline{\noalign{\ifnum0=`}\fi\hrule \@height \arrayruleHwidth
         \futurelet \@tempa\@xhline}
     \makeatother

\newcommand{\dwedge}{\,\dot\wedge\,}
\newcommand{\pprime}{\prime\prime}
\newcommand{\nn}{\nonumber \\}
\newcommand{\openk}{\Bbbk}
\newcommand{\openZ}{\mathbb{Z}}
\newcommand{\openC}{\mathbb{C}}
\newcommand{\JJF}{\mathop{\JJ}_{F}\,} 
\newcommand{\JJd}{\mathop{\JJ}_{\delta}\,\,} 

\begin{document}

\title{On the Hopf algebraic origin\\
of Wick normal-ordering} 
\author{%
Bertfried Fauser\\ 
Universit\"at Konstanz\\ 
Fachbereich Physik, Fach M678\\ 
D-78457 Konstanz\\
E-mail: Bertfried.Fauser@uni-konstanz.de}
\date{July 1, 2000}
\maketitle

\begin{abstract} 
A combinatorial formula of G.-C. Rota and J.A. Stein is taken to perform
Wick-re-ordering in quantum field theory. Wick's theorem becomes
a Hopf algebraic identity called Cliffordization. The combinatorial method
relying on Hopf algebras is highly efficient in computations and yields
closed algebraic expressions. 
\end{abstract}

\section{Introduction}

Quantum field theory needs due to the quantization
of fields an operator ordering. Using e.g. canonical quantization for
bosonic fields $b(\vec{r},t)$ and fermionic fields $\psi(\vec{r},t)$ 
and their canonical conjugates $\Pi_b$ and $\Pi_\psi$ one {\it assumes}
the canonical (anti) commutation relations:
\begin{eqnarray}
  \Big[\Pi_{b(\vec{r},t)},b(\vec{r^\prime},t)\Big]_- &=&
        \delta(\vec{r}-\vec{r^\prime}) \nn
 \Big\{\Pi_{\psi(\vec{r},t)},\psi(\vec{r^\prime},t) \Big\}_+ &=&
        \delta(\vec{r}-\vec{r^\prime}).
\end{eqnarray}
Hence one is forced to introduce ordered monomials in this variables to 
span the space of the theory. Feynmann introduced the time-ordering 
device in \cite{Fey}. However, for practical calculations one has to 
change the ordering to the so-called normal-ordering \cite{Wic,Dys}. 
This transition is motivated in terms of Feynmann diagrams as passing 
over to one-particle irreducible graphs, see e.g. \cite{ItzZub}. However,
this should be compared with the transition in thermodynamics formulated
in the BBGKY hierarchy to Ursell functions \cite{StuRie,RieStu}. From
this comparison one learns that the process of transforming a hierarchy
from time-ordered to normal-ordered correlation functions removes the
two-particle correlations in higher correlation functions. The
thermodynamic transition to Ursell functions does remove any $(n-1)$-point
correlation from $n$-point correlation functions. The well known
$\tau_n$-- and $\phi_n$-functions \cite{Lur} for time-- and
normal-ordered correlation functions are given as:
\begin{eqnarray}
\tau_n(1,\ldots,n) &:=& {}_{Phys}\langle 0 \mid
{\cal T}(\psi(\vec{r}_1,t_1),\ldots,\psi(\vec{r_n},t_n))
\mid 0 \rangle_{Phys} \nn
\phi_n(1,\ldots,n) &:=& {}_{Phys}\langle 0 \mid
{\cal N}(\psi(\vec{r}_1,t_1),\ldots,\psi(\vec{r_n},t_n))
\mid 0 \rangle_{Phys}
\end{eqnarray}
where ${}_{Phys}\langle 0\mid \ldots \mid 0 \rangle_{Phys}$ is the
expectation value w.r.t. the actual vacuum of the theory
--not a Fock vacuum in general, due to Haag's theorem \cite{Haa}--
and ${\cal T}, {\cal N}$ are time- and normal-ordering operators. The 
contraction --sometimes called covariance-- is denoted as 
${\cal C}(\psi_{I_1},\psi_{I_2})$ which could be in principle any 
scalar valued function but will be later on assigned as propagator $F$, 
see  \cite{GliJaf,Fau-thesis,Fau-trans,FauStu}. The hierarchy of $\tau$-- 
and $\phi$-correlation functions is connected via Wick's theorem as:
\begin{eqnarray}
\tau_0 &=& \langle 0 \mid 0 \rangle \,\,=\,\, 1 \nn
\tau_1(1) &=& \phi_1(1) \nn
\tau_2(12) &=& \langle 0 \mid {\cal T}(\psi_1 \psi_2) \mid 0 \rangle
\,\,=\,\, \langle 0 \mid {\cal N}(\psi_1 \psi_2) \mid 0 \rangle +
{\cal C}(\psi_1 \psi_2) \nn
&=& \phi_2(12) + {\cal C}(\psi_1 \psi_2) \nn
\tau_3(123) &=& \phi_3(123) + {\cal C}(\psi_1 \psi_2)\tau_1(3)
+{\cal C}(\psi_2 \psi_3)\tau_1(1) + {\cal C}(\psi_3 \psi_1)\tau_1(2) \\
&\vdots& \nonumber
\end{eqnarray}
We have simplified our notation and dropped the subscript {\it Phys} from
the vacuum. Furthermore we use {\it integers as variables\/} to denote the
index sets of the field operators and correlation functions, which might be
algebraic or continuous. Now, irreducibility means that an $\tau_n$-function
cannot be written as a product of $\tau_{n-r}$-functions ($r \ge 1$) e.g.
$\tau_4(1234) = \tau_2(12) \tau_2(34)$ would be reducible.

If one is interested in composite particle calculations one has to remove
lower order correlations from the desired correlation functions. This can
be achieved in a non-perturbative manner as shown in \cite{StuBor}. The
Wick re-ordering theorem can be easily proved by Clifford algebraic
methods \cite{FauAbl} unveiling its hidden geometric origin.

{\bf Remark:} Caianiello \cite{Cai} tried some time ago to connect
time-- and normal-ordered correlation functions using Clifford algebras.
Our approach is different, since we do {\it not\/} transform the Grassmann
wedge products into Clifford products as Caianiello did, but into another
{\it doted} wedge product, see below. Since the Schwinger sources force
the $\tau_n$- and $\phi_n$-functions to be antisymmetric, this cannot
be achieved using a Clifford basis without destroying the invariance under
basis transformations. Our approach is however ${\bf sl}(n)$ invariant.
As long as {\it normalized} states are considered, that is quotients of
expectation values, this approach is ${\bf gl}(n)$ invariant.

In this paper we will show, that the transition from time- to
normal-ordering is based on a Hopf algebra structure of the Schwinger
sources of quantum field theory. We introduce the well known Shuffle-Hopf
algebra \cite{Swe} which could be called Grassmann-Hopf algebra also. This
point of view is of particular use in our case. Then we use a combinatorial
formula of Rota and Stein \cite{RotSte} to show that there are infinitely
many Grassmann algebras which are {\it not\/} isomorphic as Hopf algebras.
It is shown that the {\it Rota-Stein Cliffordization\/} --employed
in a certain sense-- is exactly a closed form of the Wick transformation.
This should be compared with the cumbersome recursive process usually
performed in such reorderings, see \cite{ItzZub}. The computational 
benefits of the Hopf algebraic method compared with the usually employed
and non-perturbative method of Stumpf is discussed.

\section{Non-perturbative normal-ordering:}

\subsection{Using generating functionals}

Driven by needs of composite particle quantum field theory Stumpf introduced
the non-perturbative normal-ordering \cite{StuBor}. This method is based
on generating functionals --avoiding path integrals for certain reasons--
to describe the Schwinger-Dyson hierachy of quantum field theory. A closed
formula is then given for the transition from time- to normal-ordering and
vice versa. However, if actual computations have to be performed, a
{\it replacement\/} mechanism is effectively used. We define the (fermionic)
sources $j_K(\vec{r},t) \cong j_I \cong j_n$ using abstract indices or
even integers to denote the set of relevant quantum numbers --the
bosonic case can be handled along the same lines 
\cite{Cru,Fau-thesis,Fau-trans,FauStu}-- and their duals 
$\partial_K(\vec{r},t) \cong \partial_I \cong \partial_n$
which have to fulfil
\begin{eqnarray}
\left\{ \partial_{I_1},\partial_{I_2} \right\} \,:=\, 0 &&
\left\{ j_{I_1},j_{I_2} \right\} \,:=\, 0 \nn
\left\{ \partial_{I_1},j_{I_2} \right\} &:=& \delta_{I_1I_2}.
\end{eqnarray}
Furthermore we define the functional Fock state for convenience as
\begin{eqnarray}
\partial_I \mid 0 \rangle_F \,=\, 0&& \forall I.
\end{eqnarray}
Then we are able to write down the generating functionals which
code the hierarchy as
\begin{eqnarray}
\left\vert {\cal T}(j)\right \rangle
&:=&
\sum_{n=0}^\infty \frac{i^n}{n!} \tau_n(1,\ldots,n)
j_1 \ldots j_n \mid 0 \rangle_F
\end{eqnarray}
or
\begin{eqnarray}
\left\vert {\cal N}(j)\right \rangle
&:=&
\sum_{n=0}^\infty \frac{i^n}{n!} \phi_n(1,\ldots,n)
j_1 \ldots j_n \mid 0 \rangle_F\,.
\end{eqnarray}
The functional form of quantum field theory may be found in \cite{StuBor}
where it is shown that this formalism is able to replace usual methods
as e.g. the pathintegral approach.

Let now $F_{K_1K_2}(\vec{r}_1,t_1,\vec{r}_2,t_2) \cong F_{I_1I_2}$ be
the exact propagator of the theory. One can prove the following
theorem \cite{StuBor,FauAbl,Fau-trans}:
\begin{eqnarray}
\left\vert {\cal T}(j)\right\rangle
&=&
e^{-\frac{1}{2} F_{I_1I_2}j_{I_1}j_{I_2}}
\left\vert {\cal N}(j)\right\rangle \nn
\left\vert {\cal N}(j)\right\rangle
&=&
e^{\frac{1}{2} F_{I_1I_2}j_{I_1}j_{I_2}}
\left\vert {\cal T}(j)\right\rangle \,.
\end{eqnarray}
Expanding the series and comparing coefficients of the $j$-sources
one arrives at Wick's theorem. The factor ${1}/{2}$ was introduced
to meet the definitions of \cite{StuBor} and could be absorbed in $F$.

\subsection{Using Clifford algebras}

It was shown in a series of publications 
\cite{Fau-thesis,Fau-trans,Fau-vac,FauStu} 
that quantum field theory can be re-formulated in terms of --infinite
dimensional-- Clifford algebras of arbitrary bilinear form
--see \cite{Che,Ozi-FGTC,AblLou,Fau-mandel,Ozi-CAOM,Fau-hecke,Fau-trans}-- 
now called {\it quantum Clifford algebras\/} \cite{FauAbl}. Essentially
one performs the following steps:
\begin{itemize}
\item[i)]
Let $V=\langle j_I \rangle$ be the linear space spanned by the
Schwinger-sources $j_I$.
\item[ii)]
Build the exterior algebra --symmetric algebra for bosons-- over this
space as the formal polynomial ring in the anti-commuting sources
\begin{eqnarray}
&&\left\{ j_{I_1}, j_{I_2} \right\}_+ \,=\, 0 \nn
&&\bigwedge V \,=\, \openC \oplus V \oplus V \wedge V \oplus \ldots \nn
&&\left\vert {\cal T}(j)\right\rangle \,\in\, \bigwedge V,\quad
  \left\vert {\cal N}(j)\right\rangle \,\in\, \bigwedge V.
\end{eqnarray}
\item[iii)]
Define the space $V^*$ of linear forms on $V$ as the span of
the dual bases $\partial_I$, i.e. $V^* \,=\, <\partial_I >$ and build 
up the dual exterior algebra:
\begin{eqnarray}
V^* &=& \langle \partial_I \rangle \nn
\left\{ \partial_{I_1}, \partial_{I_2} \right\}_+ &=& 0 \nn
\left\{ \partial_{I_1}, j_{I_2} \right\}_+ &=& \delta_{I_1I_2}.
\end{eqnarray}
This space is defined to be reflexive since the same index set
is used for $j$ and $\partial$ bases. We use also the notation
$\partial_I = j_I \JJd$ where we introduced the contraction. Indeed,
this can be written basis free, e.g. $x \JJd = x_I \partial_I$ using
summation convention for discrete and continuous parts of the index
set.
\item[iv)]
Extend this setting to an action of $\bigwedge^* V \cong
(\bigwedge V)^*$ on $\bigwedge V$ in the following way
\cite{Che,Ozi-FGTC,Lou,Fau-trans,Fau-hecke,FauAbl}:
\begin{eqnarray}
\label{schwinger}
i)   && \partial_{I_1}(j_{I_2}) \,=\, \delta_{I_1I_2} \nn
ii)  && \partial_I (AB) \,=\, (\partial_I A)B +
        \hat{A}(\partial_I B) \nn
iii) && A^* (BC) \,=\, A^* (B^* C) \,.
\end{eqnarray}
The notation is as follows: $A,B,C \,\in\, V$, i.e.
$A  = A_{I_1,\ldots,I_n} j_{I_1} \ldots j_{I_n}$,
$A^*, B^* \,\in\, V^*$, $\hat{A} = (-1)^r A_{I_1,\ldots,I_r}
j_{I_1}\ldots j_{I_r}$ and, note the reversion of indices here 
$A \mapsto A^* = A_{I_1, \ldots I_n} 
\partial_{I_n} \ldots \partial_{I_1}$. 
This can be recast entirely --avoiding wedge and contraction--
in Clifford algebraic form by index doubling, see \cite{FauAbl,Fau-trans}.
\item[v)]
Define the field operators as Clifford algebra elements obtained by
the Clifford map according to Chevalley deformation:
\begin{eqnarray}
\psi_K(\vec{r},t) \,\equiv\, \psi_I &:=&
\partial_I + B_{I_1I_2}j_{I_2} \nn
&\cong& \partial_K(\vec{r},t)
   + B_{K_1K_2}(\vec{r}_1,t_1,\vec{r}_2,t_2) j_{K_2}(\vec{r}_2,t_2)
\end{eqnarray}
where summation and integration is once again implicit. This is a 
particular form of a deformation quantization.
\end{itemize}

Define now explicitely the wedge product denoted by $\wedge$ as
the sign for the product of the Schwinger sources. Furthermore,
we use index free notation now to shorten the formulas. Let us
furthermore define the `bi`-vector
$F := F_{I_1I_2} j_{I_1} \wedge j_{I_2}$. Let us introduce a
{\it second\/} wedge product, the doted-wedge \cite{LouRie} denoted 
by $\dwedge$, defining on $x,y \in V$
\begin{eqnarray}
x \dwedge y &:=& x \wedge y + F \JJd (x \wedge y) \nn
        &=& x \wedge y + F_{xy}.
\end{eqnarray}
Observe that $x \dwedge y = -y \dwedge x$ and $\dwedge$ is indeed
antisymmetric and a proper exterior product which can be extended
to the whole algebra $\dot{\bigwedge} V$.

In Ref. \cite{FauAbl} the following theorem was proved:

Let $e^F$ denote the exterior exponential of $F\,\in \bigwedge^2 V$
and $A=A(j,\partial)$ an arbitrary operator in $End(\bigwedge V)$,
then we have:
\begin{eqnarray}
e^{F} \wedge A(j,\partial) \wedge e^{-F} &=& A^{\dwedge}(j,\partial)
\,=\, A^{\wedge}(j,d) \nn
d &:=& \partial - Fj \label{def-d}
\end{eqnarray}
for operators and 
\begin{eqnarray}
e^{F} \mid {\cal T}(j)\rangle^{\wedge} &=& \mid {\cal N}(j) 
\rangle^{\dwedge} \nn
e^{-F} \mid {\cal N}(j)\rangle^{\dwedge} &=& \mid {\cal T}(j) 
\rangle^{\wedge} 
\end{eqnarray}
for functional states. The peculiar feature of this transition is, that
in a first step {\it only\/} the product is transformed, hence we have:
\begin{eqnarray}
\mid {\cal N}(j) \rangle^{\dwedge} &=&
\sum \frac{i^n}{n!} \tau_n(1,\ldots,n) 
j_{I_1} \dwedge \ldots \dwedge j_{I_n} \mid 0 \rangle_F.  
\end{eqnarray}
The {\it normal-ordered\/} functional is thus written with the
{\it time-ordered} correlation functions. The usually obtained
normal-ordered correlation functions appear only after we have
{\it re-written\/} the normal-ordered functional in terms of the
{\it old wedge\/} $\wedge$. We arrive at
\begin{eqnarray}
\mid {\cal N}(j) \rangle^{\wedge} &=&
\sum \frac{i^n}{n!} \phi_n(1,\ldots,n) 
j_{I_1} \wedge \ldots \wedge j_{I_n} \mid 0 \rangle_F  
\end{eqnarray}
where the $\phi_n$-functions are connected to the $\tau_n$-functions
via the Wick theorem w.r.t. the contraction --or covariance-- $F$.
This transition is called Wick isomorphism $\phi$ 
and is a $\openZ_2$-graded algebra homomorphism \cite{FauAbl}.

Symbolically we write this as
\begin{eqnarray}
\cl(V,Q) &=& \phi \circ \cl(V,B),
\end{eqnarray}
where $\cl(V,Q)$ and $\cl(V,B)$ are Clifford algebras over the space
$V$ w.r.t. the quadratic form $Q$ or the --not necessary symmetric--
bilinear form $B$, see \cite{Fau-trans,Fau-hecke,FauAbl}. In the 
present case, we look at both `Clifford algebras` as degenerated 
ones, i.e. as Grassmann algebras, that is we set $Q\equiv 0$ and 
let $B = -B^T$ be a totally antisymmetric form, which is also 
degenerated since $1/2(B+B^T) \equiv 0$.

{\bf Example 1:} [Schwinger-Dyson hierarchy for a free Hamiltonian]
Let $H$ be the functional Hamiltonian of a free fermionic field.
Such an $H$ takes the form
\begin{eqnarray}
H^\wedge(j,\partial) &:=& D_{I_1I_2} j_{I_1}\partial_{I_2},
\end{eqnarray}
where $D_{I_1I_2}$ is the kinetic operator e.g. a d`Alembertian or
Laplacian, see \cite{StuFauPfi,FauStu,Fau-trans} for a detailed model 
of spinor QED. The normal-ordered operator is obtained as
\begin{eqnarray}
H^{\dwedge}(j,\partial) &=& e^{F} \wedge H^{\wedge}(j,\partial)
\wedge e^{-F} \nn
&=& H^{\wedge}(j,d) \nn
&=& D_{I_1I_2}j_{I_1}(\partial_{I_2} - F_{I_2I_3}j_{I_3}) \nn
&=&  D_{I_1I_2}j_{I_1}\partial_{I_2} - D_{I_1I_2}F_{I_2I_3}j_{I_1}j_{I_3}
\end{eqnarray}
and the generating functional of the Schwinger-Dyson hierachy
transforms as follows:
\begin{eqnarray}
e^{F} \wedge H^{\wedge}(j,\partial) \mid {\cal T}(j) \rangle>^{\wedge}
&=& e^{F} \wedge E \mid {\cal T}(j)\rangle^{\wedge} \nn
H^{\dwedge}(j,\partial) \mid {\cal N}(j) \rangle>^{\dwedge}
&=& E \mid {\cal N}(j)\rangle^{\dwedge} \nn
H^{\wedge}(j,d) \mid 
{\cal N}(j) \rangle>^{\wedge}
&=& E \mid {\cal N}(j)\rangle^{\wedge}
\end{eqnarray}
This example shows, that finally the transition from time- to 
normal-ordering is given by re-expressing the doted-wedge $\dwedge$ 
in terms of the undoted wedge $\wedge$ and vice versa. This can be 
achieved by the above given formal substitution in the operators e.g. 
$H^{\dwedge}(j,\partial) = H^{\wedge}(j,d)$, using (\ref{def-d}) and 
expanding the doted wedges in the generating functional, see 
\cite{Fau-thesis,Fau-trans}. However, the deeper origin of the need of 
a normal-ordering remains hidden. The root of such a re-ordering will 
be found in the Hopf algebraic structure of the Grassmann algebra.

\section{Grassmann-Hopf algebra}

\begin{figure}
\centerline{\epsfig{figure=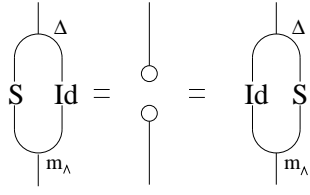,width=4truecm}}
\caption{Tangle definition of the antipode axioms.}\label{fig-1}
\end{figure}

We have already introduced the Grassmann algebra above. To turn it into
a Hopf algebra, we have to add in a compatible way a co-algebra structure
and an antipode. Let us denote the unit map which injects the real or 
complex field into the algebra with $\eta$, $\eta : \openk \mapsto 
\bigwedge V$, while $\wedge \equiv m_\wedge$, 
$m_\wedge : \bigwedge V \otimes \bigwedge V \mapsto \bigwedge V$ 
is the product map. Let us furthermore denote 
the linear space underlying the Grassmann algebra as $W$, thus 
$W = \langle \bigwedge V \rangle$, then we can describe the Grassmann 
algebra by the triple $\bigwedge V = (W,m_\wedge,\eta)$. The co-algebra 
structure is then given by a {\it diagonalization\/} $\Delta$ 
--called also co-product-- and a co-unit $\epsilon$ which arise naturally 
from 'dualizing' the algebra structure in a functorial sense, see 
\cite{Swe,Maj}. The compatibility of algebra and co-algebra structure 
requires the diagonalization and co-unit to be algebra homomorphisms. 
That is we require that:
\begin{eqnarray}
\epsilon(Id) &=& 1 \nn
\epsilon(j_I) &=& 0 \nn
\epsilon(A \wedge B) &=& \epsilon(A) \wedge \epsilon(B).
\end{eqnarray}
While the co-product has to fulfil:
\begin{eqnarray}
\Delta(Id) &=& Id \otimes 	d \nn
\Delta(j_I) &=& j_I \otimes Id + Id \otimes j_I \nn
\Delta(A \wedge B) &=& \Delta(A) \wedge \Delta(B)
\end{eqnarray}
obeying an $\openZ_2$-graded tensorproduct. We introduce for further
usage also the Sweedler notation of co-products as \cite{Swe}
\begin{eqnarray}
\Delta(x) &=& \sum_x x_{(1)} \otimes x_{(2)},
\end{eqnarray}
where we omit the subscript at the sum-sign or even the sum-sign itself
which is them implicit. The above written wedge-product for
{\it tensors\/} 
--$(a\otimes \ldots \otimes b) \wedge (c \otimes \ldots \otimes d)$-- 
needs for its evaluation the graded switch $\tau_\wedge$ which is defined 
as ($u \in \bigwedge_r V$, $ v \in \bigwedge_s V$):
\begin{eqnarray}
\tau_\wedge &:& \bigwedge_r V \otimes \bigwedge_s V \mapsto 
        \bigwedge_s V \otimes \bigwedge_r V \nn
 && \tau_\wedge ( u \otimes v) \,=\, (-1)^{rs} (v \otimes u),
\end{eqnarray}
extended by linearity to all elements of $\otimes^2 W$. Finally we 
define the antipode $S^\wedge\equiv S \in End(W)$
--i.e. the convolutive inverse of $Id$-- as:
\begin{eqnarray}
S^\wedge &:& \bigwedge_r V \mapsto \bigwedge_r V \nn
 && S^\wedge(x) \,=\, (-1)^r x,
\end{eqnarray}
also extended by linearity. Note that this is exactly the grade
involution of the Grassmann algebra. The sextuple
\begin{eqnarray}
H_\wedge &:=& (W,m_\wedge,\eta,\Delta,\epsilon,S^\wedge)
\end{eqnarray}
is the Grassmann-Hopf algebra, which is in a certain sense unique 
(universality property) see \cite{Swe}. It can be checked that with the 
above given definitions all axioms for a Hopf algebra are met, 
especially the antipode axioms are fulfilled [see Figure \ref{fig-1}]:
\begin{eqnarray}
i)   && \wedge \circ (S \otimes Id) \circ \Delta 
\,=\, \eta \circ \epsilon \nn
ii)  && \wedge \circ (Id \otimes S) \circ \Delta 
\,=\, \eta \circ \epsilon.
\end{eqnarray}
Observe, that in our case $S^2 = Id$ and the Grassmann-Hopf
algebra
is $\openZ_2$-graded co-commutative.

\section{The Rota-Stein Cliffordization formula as Wick theorem}

In Ref. \cite{RotSte} Rota and Stein introduced a deformed product on 
Hopf algebras. Adding the structure of a Laplace pairing --definitions 
see below-- one can add to the universal structure of a Grassmann-Hopf
algebra a new product which relays on the Laplacian pairing. This 
corresponds to an algebra deformation of the Grassmann algebra into 
a Clifford algebra. We will, however, use this mechanism to connect 
Grassmann-Hopf algebras which are {\it different}, i.e. non-isomorphic,
as $\openZ_n$-graded (Hopf) algebras. This can be done using a Laplace 
pairing w.r.t. an antisymmetric form, which will be afterwards defined 
to be the propagator of the theory.

We define the Laplace pairing following Ref. \cite{RotSte}. Let 
$(w\vert w^\prime)$ be a bilinear mapping from elements $w,w^\prime$ of
the Grassmann Hopf algebra $H_\wedge$ into $H_\wedge$ --we will need 
$\openk$ as a target only-- which satisfies the Laplace identities:
\begin{eqnarray}
\label{laplace}
i)   && (w \wedge w^\prime \vert w^{\pprime}) \,=\,
        \sum \pm (w \vert w^{\pprime}_{(1)}) \wedge
                 (w \vert w^{\pprime}_{(2)})
\nn 
ii)  && (w \vert w^\prime \wedge w^{\pprime}) \,=\,
        \sum \pm (w_{(1)} \vert w^\prime) \wedge
                 (w_{(2)} \vert w^{\pprime})
\\
iii) && 
(w \vert w^\prime)_{(1)} \wedge ((w \vert w^\prime)_{(2)} \vert w^{\pprime})
\,=\,
\sum \pm 
(w_{(1)} \vert w^\prime_{(1)}) \wedge ((w_{(2)} \vert w^\prime_{(2)}) 
\vert w^{\pprime})
\nn
iv)  &&
(w \vert (w^\prime \vert w^{\pprime})_{(1)}) \wedge  
(w^\prime \vert w^{\pprime})_{(2)} 
\,=\,
\sum \pm 
(w \vert (w^\prime_{(1)} \vert w^{\pprime}_{(1)})) \wedge 
        (w^\prime_{(2)} \vert w^{\pprime}_{(2)}). \nonumber
\end{eqnarray} 
The signs $\pm$ have to be chosen due to the action of the graded switch
to produce the correct permutations. If the the bilinear form is scalar 
valued, i.e. in $\openk$ as we assume, the wedge products can be safely 
removed. Relations $i)$ and $ii)$ are the Hopf algebraic expression of 
the Laplace-expansion of determinants. Relations $iii)$ and $iv)$ state 
the compatibility of the bilinear form and the co-product being an 
algebra homomorphism. 

\begin{figure}
\centerline{\epsfig{figure=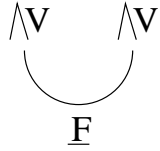,width=2truecm}}
\caption{Tangle for the scalar valued pairing.}\label{fig-2}
\end{figure}

We use tangles \cite{Yet} to make some of the relations more feasible. 
The tangle for the scalar valued pairing is given as in Figure 
\ref{fig-2}.

\begin{figure}
\centerline{\epsfig{figure=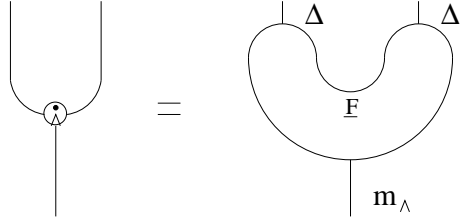,width=6truecm}}
\caption{Tangle definition of the `Rota-sausage'.}\label{fig-3}
\end{figure}

We are now ready to define with Rota and Stein the deformed product as:
\begin{eqnarray}
\label{sausage}
w \dwedge w^\prime &:=& 
\sum \pm w_{(1)}
\wedge ( w_{(2)} \vert w^\prime_{(1)} )_F \wedge w^\prime_{(2)}
\end{eqnarray}
or in terms of tangles [see Figure \ref{fig-3}] as the so called
`Rota-sausage` \cite{Ozi-ixtapa}. Note, that the deformed product
is given by a {\it non-local\/} graph. This process is called
{\it Cliffordization\/} since it is used usually to introduce a
non-trivial symmetric bilinear form. The non-locality of the
`sausage` might have implications on Feynmann diagrams build
up from such products. Indeed we found in Ref. \cite{Fau-vertex}
that some singularities which arise due to re-ordering in
vertices of non-linear spinor field models dissappear due to
this rearrangements. Hopf algebras are recently used successfully 
in perturbative renormalization theory \cite{Con,Kre,RosVer}.

Observe the perplexing fact, that this definition of the doted wedge 
product is generic for {\it any\/} element of the underlying space 
$W = <\bigwedge V>$ of Grassmann polynomials. This contrasts strongly 
the {\it recursive\/} definition of the doted wedge product due to 
Chevalley deformation of the Grassmann wedge product \cite{Che,Che2}. 
There one has ($x \in V$)
\begin{eqnarray}
\gamma_x &:& \bigwedge V \,\mapsto\, \bigwedge V \nn
x &\mapsto& \gamma_x \,:=\, x \JJF\,\,\, +\, x\, \wedge   
\end{eqnarray}
which works {\it only\/} for $x \in V$ and has to be extended by the 
rules given above for the Schwinger sources (\ref{schwinger}). At 
this place we note that the grade involution appearing in 
(\ref{schwinger}-ii) is exactly the antipode $S^\wedge$.

The only weak point in our definition is, that we have not yet given 
a computational definition of the pairing. The pairing can be axiomatized 
\cite{DouRotStei}. However, the most important relations are the two
Laplace expansions (\ref{laplace}-i) and (\ref{laplace}-ii) which allow
us by applying the coproduct to decompose the pairing into pairings
which contain only elements of the space $V$. The Laplace pairing is 
exactly a pairing which allows such a decomposition.

However, we will give a second definition of the pairing, which is
equivalent to the above one, but might be more familiar to physicists. 
We want nevertheless to stress that the evaluation of this expression is
done by applying the Laplace expansion rules given above, beside the fact 
that this time the Hopf algebraic nature of the expansion is disguised.

Let $w_r = x_1 \wedge \ldots \wedge x_r$ and 
$w^\prime_s = y^\prime_s \wedge \ldots \wedge y^\prime_1$ 
with $x_i,y_j \in V$ and define
\begin{eqnarray}
(w_r \vert w^\prime_s )_F &:=&
\left\{
\begin{array}{ll}
{\rm det}(x_i \vert y^\prime_j )_F & r = s \\
0 & {\rm otherwise }
\end{array}
\right. 
\end{eqnarray}
and extend it by linearity to the whole space $W$. This can be rewritten
using the projection onto the scalar part 
$\langle . \rangle^\wedge_0 \,:\, W \,\mapsto\, \openk$ 
and the contraction $\JJF$ w.r.t. $F$ as:
\begin{eqnarray}
( w \vert w^\prime ) &=& \langle w \JJF w^\prime \rangle^\wedge_0 \nn
		    &=& \epsilon( w \JJF w^\prime ).
\end{eqnarray}
We note, that the projection onto the scalars
$\langle \,.\, \rangle^\wedge_0$ is exactly the co-unit $\epsilon$
by definition.

The following example shows how this mechanism works. They have been
produced for higher dimensions using the computer algebra packages
CLIFFORD for Mapel, developed by Rafa{\l} Ab{\l}amowicz \cite{Abl,CLI}
and BIGEBRA \cite{BIG}. However, the below given example can still
done easily by hand.

{\bf Example 2:} [Hopf algebraic Wick re-ordering] 
Let $x,y \in V$, let $\circ$ denote the concatenation of operations 
and calculate the `sausage`, i.e. formula (\ref{sausage}):
\begin{eqnarray}
x \,\dwedge\, y &=& \wedge \otimes \wedge
\circ ( Id \otimes (\, .\, \vert \, .\,) \otimes Id)
\circ (\Delta \otimes \Delta) ( x \otimes y) \nn
&=& \wedge \otimes \wedge
\circ ( Id \otimes (\, .\, \vert \, .\,) \otimes Id)
(x \otimes Id \otimes y \otimes Id \nn
&&
+x \otimes Id \otimes Id \otimes y
+Id \otimes x \otimes y \otimes Id
+Id \otimes x \otimes Id \otimes y) \nn
&=& \wedge \otimes \wedge
(0 + (Id \vert Id)_F x \otimes y + (x \vert y)_F Id \otimes Id +0) \nn
&=&  x \wedge y + F_{x,y} Id.  
\end{eqnarray}
Thus we obtain as the worked out Rota-sausage the relation (\ref{sausage}),
of the doted wedge, expressed in undoted wedges. Let us compute
a product of three doted wedges ($x_i \in V$)
\begin{eqnarray}
x_1 \dwedge x_2 \dwedge x_3 &=&
x_1 \dwedge (F_{23} Id + x_2 \wedge x_3) \nn
&=&
x_1 \wedge x_2 \wedge x_3 + x_1 F_{23} + x_2 F_{31} +x_3 F_{12}. 
\end{eqnarray}
It seems to be that we have to resolve the tangle recursively, but this
is for demonstration only, one can write down a closed formula using
nested Rota-sausages to obtain the result directly. In fact, we see
clearly how the $\tau_n$- and $\phi_n$-correlation functions are
interrelated due to this expansion. As a last calculation, we find
the doted wedge product of two elements of $x_1 \wedge x_2 $,
and $x_3 \wedge x_4$ both $\in \wedge^2 V$
\begin{eqnarray}
\lefteqn{(x_1 \wedge x_2) \dwedge (x_3 \wedge x_4)} && \nn
&=& (\wedge \otimes \wedge)
\circ ( Id \otimes (\, .\, \vert \, .\,) \otimes Id)
\circ (\Delta \otimes \Delta) ( (x_1 \wedge x_2) \otimes (x_3 \wedge x_4)) \nn
&=& (\wedge \otimes \wedge)
\circ ( Id \otimes (\, .\, \vert \, .\,) \otimes Id)
\big(
(x_1 \wedge x_2) \otimes Id \otimes (x_3 \wedge x_4) \otimes Id \nn
&&
+(x_1 \wedge x_2) \otimes Id \otimes x_3 \otimes x_4
-(x_1 \wedge x_2) \otimes Id \otimes x_4 \otimes x_3 \nn
&&
+(x_1 \wedge x_2) \otimes Id \otimes Id \otimes (x_3 \wedge x_4)
+x_1 \otimes x_2 \otimes (x_3 \wedge x_4) \otimes Id \nn
&&
+x_1 \otimes x_2 \otimes x_3 \otimes x_4
-x_1 \otimes x_2 \otimes x_4 \otimes x_3 \nn
&&
+x_1 \otimes x_2 \otimes Id \otimes (x_3 \wedge x_4)
-x_2 \otimes x_1 \otimes (x_3 \wedge x_4) \otimes Id \nn
&&
-x_2 \otimes x_1 \otimes x_3 \otimes x_4
+x_2 \otimes x_1 \otimes x_4 \otimes x_3 \nn
&&
-x_2 \otimes x_1 \otimes Id \otimes (x_3 \wedge x_4)
+Id \otimes (x_1 \wedge x_2) \otimes (x_3 \wedge x_4) \otimes Id \nn
&&
+Id \otimes (x_1 \wedge x_2) \otimes x_3 \otimes x_4
-Id \otimes (x_1 \wedge x_2) \otimes x_4 \otimes x_3 \nn
&&
+Id \otimes (x_1 \wedge x_2) \otimes Id \otimes (x_3 \wedge x_4)
\Big) \nn
&=& (Id \vert Id)_F x_1 \wedge x_2 \wedge x_3 \wedge x_4
\nn &&
+(x_2 \vert x_3)_F x_1 \wedge x_2
-(x_2 \vert x_4)_F x_1 \wedge x_3 \nn
&&
-(x_1 \vert x_3)_F x_2 \wedge x_4
+(x_1 \vert x_4)_F x_2 \wedge x_3
\nn
&& (x_1 \wedge x_2 \vert x_3 \wedge x_4)_F Id.
\end{eqnarray}

It is obvious, that one can define an inverse mapping
in the same fashion as long as the pairing is non-degenerate.
This allows one to expand the undoted wedges into the doted ones
using the Rota-Stein Cliffordization w.r.t. the bilinear form $-F$.
This is equivalent to the expansion of the normal-ordered 
$\phi_n$-correlation functions in terms of the time-ordered 
$\tau_n$-correlation functions.

{\bf Remark:} In spite of the fact, that the Grassmann algebras w.r.t.
the two wedges --doted and undoted-- are isomorphic as $\openZ_n$-graded
algebras, this is {\it not\/} the case for the Hopf algebras $H_\wedge$
and $H_{\dwedge}$. This can be seen easily by evaluating the co-unit
$\epsilon_\wedge$ on a doted and undoted wedge ($x,y \in V$):
\begin{eqnarray}
\epsilon_\wedge(Id) &=& 1 \,=\, \epsilon_{\dwedge}(Id) \nn
\epsilon_\wedge(x) &=& 0 \,=\, \epsilon_{\dwedge}(x) \nn
\epsilon_\wedge(x \wedge y) &=& 0 \nn
\epsilon_\wedge(x \dwedge y)
&=& \epsilon_\wedge(x \wedge y + F_{x,y} Id )
\,=\, F_{x,y} \,\not=\, 0\,.
\end{eqnarray}
Obviously we find that $\epsilon_\wedge$ is {\it not\/} the co-unit of
$H_{\dwedge}$ and that we are forced to introduce a new co-unit
$\epsilon_{\dwedge}$ there. An analogous situation is found for the
co-products, which are also not identical. This opens the interesting
possibility to have a family of co-units and co-products for an --up
to isomorphy-- given Grassmann algebra to make it into a Grassmann-Hopf
algebra. In Ref. \cite{Fau-vac} we showed, how such a process can be
used to introduce different vacua and even condensation phenomena.
We learn from our present work, that the term `vacuum` is connected
with the co-unit of a Hopf algebra. In fact the co-unit is a sort of
expectation value of the algebra elements which constitute the
operators. The Hopf algebraic non-isomorphic $\openZ_n$-gradings
are responsible for this feature. 

We conjecture furthermore, that this process is involved in the recent
development of A. Connes and D. Kreimer \cite{Con,Kre,RosVer} of a theory
on perturbative renomalization of quantum field theory, where the
antipode generates all the counterterms in the forest theorems.
However, their theory uses the $\openZ_2$-grading only and is thus
not sensible to the finer $\openZ_n$-grading used in our work.

\section{Conclusion}

We showed that the process of Wick-reordering is governed by the
Grassmann-Hopf algebra structure uniquely assigned to Schwinger
sources and a definite wedge product. Introducing a doted wedge product,
using Rota-Stein Cliffordization w.r.t. an antisymmetric bilinear form
$F$ and the therefrom induced pairing, we found a closed formula
for re-ordering time-- into normal-ordered and normal-- into
time-ordered $n$-point correlation functions. The Hopf algebraic nature
of this process was exhibited and its importance in quantum field theory
was demonstrated. It was demonstrated, that the $\openZ_n$-grading of a
Grassmann-Hopf algebra is {\it not} preserved under such an transition.
This was connected to different vacua underlying the particular theory.
The Stumpf approach to non-perturbative normal-ordering was given and 
compared to the Hopf algebraic method.

\end{document}